\documentstyle[12pt]{article}
\def\beq{\begin{equation}}
\def\eeq{\end{equation}}

\def\bea{\begin{eqnarray}}
\def\eea{\end{eqnarray}}
\def\ci{\cite}
\def\la{\label}
\def\le{\left}
\def\ri{\right}

\begin{document}

\begin{flushright}
hep-ph/9511470 \\
IFUNAM-FT-95-79 \\
IOA-325-95 \\
\end{flushright}
\vspace{15mm}

\begin{center}
{\large \bf INFLATION IN S-DUAL SUPERSTRING MODELS
} \\
\end{center}

\vspace*{0.7cm}

\begin{center}
{\bf A. De la Macorra$^{a}$\footnote{This work was partially
supported by CONACYT Proy. 4964-E} and S. Lola$^{b}$}
\end{center}

\vspace*{0.1cm}
\begin{center}
\begin{tabular}{c}
$^{a}${\small Instituto de Fisica, UNAM}\\
{\small Apdo. Postal 20-364,
01000  Mexico
D.F., Mexico}\\
$^{b}${\small Department of Theoretical Physics,
University of Ioannina} \\
{\small 451 10 Ioannina, Greece} \\
\end{tabular}
\end{center}

\vspace{1 cm}

\begin{center}
{\bf ABSTRACT}
\end{center}

{\small We study inflationary potentials in the framework of
 superstring theories.
Successful inflation may occur due to chiral fields, but only
after the dilaton and moduli
are stabilized. This is achieved by demanding an
S-duality invariant potential. Then,
it is possible to have inflation at any scale and even
to have more
than one  stages of inflation. This occurs for
a limited class of scalar potentials, where certain conditions
are obeyed.
Density fluctuations of $10^{-5}$ require  the  inflationary
potential to be at a scale of
$V^{1/4}\simeq 5 \times 10^{15}$ GeV.}

\noindent
\rule[.1in]{16.5cm}{.002in}

\thispagestyle{empty}

\setcounter{page}{0}
\vfill\eject

Inflationary scenarios \cite{inflation}
address several of
the problems of the standard hot big-bang theory,
such as the flatness and horizon
problems, the overabundance of topological
defects that result from the breaking
of a GUT symmetry and the origin of
density fluctuations in the universe
\cite{faults}.
Among the several possibilities,
one of particular interest is chaotic
inflation, which provides
a natural set of initial conditions for the
inflaton field \cite{chao}.
In this scheme however, in order to obtain
sufficient growth of the scale factor
as well as the correct magnitude
of density fluctuations,
the introduction of an arbitrary, tiny coupling
constant, associated with the
interaction strength in the scalar sector is required.
This is avoided in theories where
a transition from a higher dimensional to a four-dimensional
universe occurs \cite{wett}, since they are
governed by only one scale, the Planck
mass $M_{Planck}$.
The best candidate of a higher dimensional
unification is superstring theory
and there are several proposals on inflation
from superstrings \ci{stew}-\ci{am}.

In string potentials, inflation may occur only after
freezing the dilaton field,
and  before  this happens
no other field may be used as the inflaton.
Indeed, in schemes where
inflation is due to the scalar potential, it is not possible
to obtain enough e-folds of inflation while the dilaton
is a dynamical field
\cite{stein}, even when the loop effects from the 4-Fermi
gaugino interactions are included \cite{am}.
(In an alternative solution,
inflation occurs due  to the
kinetic energy  of the inflaton \ci{Venez}).
The stabilization of the dilaton field has many important
phenomenological consequences. For example, it fixes the gauge
coupling constant at the string scale and sets the hierarchy
between the masses of the fermions and their supersymmetric partners.
Even though the vev (vacuum expectation value)
of the dilaton is essential in understanding
the low energy models of string theory, there is still no definite
answer as to how it arises. A
plausible way to fix the vev of the dilaton is via
gaugino condensates \cite{gaug},\cite{axel}, which are
expected to form if the string model  has
an asymptotically free gauge group. In \cite{am} we
studied the inflationary potentials in the presence
of gaugino condensates.  Even though enough inflation can be obtained
to solve
the flatness and horizon problems, the scale at which the dilaton
freezes is in these schemes $10^{12-13}$ GeV\footnote{This scale is
obtained  by demanding that the supersymmetric
 partners of the fermions  have  masses around $1$ TeV, as required
by supersymmetry breaking arguments.}. The  density fluctuations
generated by inflation at or below the condensation scale are then
much smaller \cite{am}
than the observed by COBE \cite{COBE}.

Here, we will study
the possibility of having inflationary potentials above the
supersymmetry breaking scale, but still demanding  a stable dilaton
field. An interesting possibility to stabilize the
dilaton at a higher scale
appears  when an S-duality symmetry of the lagrangian
(conjectured by Ibanez
et al. in the framework of superstring theories)
is considered
\cite{Sdual}.
A great deal of work has been invested in studying  this symmetry
and related ones, due to the results of Seiberg and Witten
\cite{seibwit}, which were obtained for
$N=2$ SUSY. It  is still not
clear whether S-duality will survive in an $N=1$ SUSY theory.
However, we think that it is interesting to study the
consequences for inflation,
assuming that such a symmetry remains valid in $N=1$ SUSY
lagrangians.  Even more so, since there are no other alternatives to
fix the dilaton field at a large  scale of $10^{15} GeV$ that we know
of.
Demanding an S-dual potential, inflation may occur at a higher scale
than the condensation scale,
resulting to larger density fluctuations.
We will show that an S-dual potential allows  to have inflation at
any scale and also opens the possibility
of having more than one stages of
inflation in string models. Normalization of the density
fluctuations with respect to COBE  requires  a  spontaneously broken
symmetry  at around $V^{1/4}_{Inf}\simeq 5\times 10^{15} GeV$.  The
breaking of symmetries  is expected
in string models since  the structure is very rich \cite{gr}
and one expects, in general, scalar fields to acquire a vev
either at tree level
or via radiative corrections as in the breaking of the
electroweak symmetry.

To see how this works, we impose
S-duality to the  low energy superstring Lagrangian.
The simplest realization is to take  S-duality as an SL(2,Z) symmetry
generated by $S \rightarrow 1/S$ and $S \rightarrow S + i$ (this last
symmetry  is  already present in all 4-D string vacua).  In order to
achieve the SL(2,Z)  invariance one can distinguish two different
cases \ci{nilles}: (1) $f  \rightarrow f$ or (2) $f \rightarrow 1/f$,
where $f$ is the gauge kinetic function. In the first case the gauge
coupling constant,  given by $g^2=Re f$, remains invariant and the
gauge chiral superfield $W_{\alpha}$ coupled to $f$ by a term $(f
W_{\alpha}W^{\alpha})_F + h.c$ will also be invariant. On the other
hand,
 in case (2) one has $g^2 \rightarrow 1/g^2$ and the gauge chiral
superfield is no longer invariant and  one  requires additional
fields called "magnetic mesons" \ci{Sdual}
to fully realized the symmetry. It is not clear whether an N=1
supergravity
is reliable  in this case.  The gauge kinetic function can be written
in case (1) as $f=(1/2\pi) ln(j(S)-744)$ \ci{nilles}, where $j(S)$ is
the
generator of modular invariant functions\footnote{For large $S$ one
has $j(S)=1/q+744+196884\;q +O(q^2),\; q=e^{-2\pi S}$.}. This $f$
reproduces the large $S$  limit calculated perturbatively, i.e.
$f=S$. We are interested  in the scalar sector only and will
therefore not consider the gauge sector of the theory.
Since the gauge coupling constant enters in the scalar
sector via the $D$-term
with $V= \frac{1}{2 Re\,f} D^2$ then for $D=0$  the
gauge and scalar sectors decouple.

The effective $ d=4 $
superstring  model
is given by an $N = 1$
supergravity  theory \cite{r15}
with at least  four gauge singlet fields
$S$ and $T_i,  \,i=1,2,3$ as well as an
unspecified number of gauge
chiral matter superfields. The tree level scalar potential
is given by \ci{r15}
\bea
V&=&  e^{K} H + \frac{1}{2 Re f } D^2
\\
H &=& (G_{a} (K^{-1})^{a}_{b} G^{b} -3|W|^{2})
\nonumber\\
D&=& \hat{g} K^i T_i^j \phi_j + \xi
\label{eq1}\eea
where  $D$ is the auxiliary field of the  gauge vector superfield,
$T_i^j$ the generators of the gauge group, $\hat{g}$ the charge,
$\xi$ a Fayet-Illiopoulos term and
\begin{eqnarray}
G & \equiv &
K+ ln |W|^2 \nonumber \\
K & = & - \log(S_{r}) -
\log \left[ (T_{r} -\phi\bar{\phi})^3 - B \bar{B} -
T_{r} (C\bar{C}) \right ]
\label{eq2}\\
W & = &   \Omega(S) P(T,\phi,B,C) \nonumber \\
\Omega(S) & = & \eta^{-2}(S)\nonumber
\end{eqnarray}
Here $G$ is the K\"{a}hler
potential,
$T_{r} =   T + \bar{T}, S_{r}  =  S + \bar{S} $ and the indices
$a,b$ run
over all fields, i.e. the dilaton $S$, the moduli $T$ and
chiral fields
$\phi$, $B$ and $C$. The $\phi$ fields correspond to untwisted chiral
fields
while $B$ and $C$ are twisted fields. The indices $a,b$
of the functions $K$ and $W$ denote derivatives with respect to
chiral fields \cite{effective}.
All the fields are
normalized with respect to the reduced Planck mass
 $m_{p} =  M_{p}/\sqrt{8 \pi}$.
The form of $K$ is derived by a perturbative
expansion and is
valid if the arguments inside the logarithms are positive.
The term $\Omega(S)$ (where $\eta$ is the Dedekind-eta function with
modular weight 1/2) arises due to S-duality in order to cancel the
transformation of K and to render $G=K+ln|W|^2 $ invariant.
A similar situation appears for the moduli fields $T_i$. The function
$
\Omega(S)$ may be multiplied by a modular invariant function but, for
simplicity,  we have chosen it to be unity.
The chiral superpotential $P$ is a polynomial function of the chiral
fields and in particular, it contains the trilinear (Yukawa)
interactions of the
chiral
fields. Here, we work with the effective
low energy superstring Lagrangian in the Einstein
and not in the Brans-Dicke frame \ci{Bellido},
which would seem more natural in the context
of string inflation. This is
for continuity, since
most of the work in determining the  non-perturbative contributions
of $S$ and $T$ to the potential as well as
the study of the duality symmetries  for $T$ and
$S$,  has been carried out in the  Einstein frame.

{}From now on we work with the untwisted fields $\phi$,
which we found in \cite{am} to be best suited for inflation
than $B$ and $C$,
due to the form of  $K$.
Considering only untwisted fields in the sector  related
to $T_1$, the superpotential can be written as
$W=\eta^{-2}(S)\eta^{-2}(T_2)\eta^{-2}(T_3)P(T_1,\phi)$.
Here we take $P$ in the form
$P(T_1,\phi)= \lambda(T_1)P(\phi)$.
The derivatives of the K\"{a}hler potential are
\bea
G_{S} (K^{-1})^{S}_{S} G^{S}&=&|W|^2 \frac{S^2_r}{4\pi^2}|\hat
G_2(S)|^2
\nonumber\\
G_{T_i} (K^{-1})^{T_i}_{T_i} G^{T_i}&=&|W|^2 \frac{
T^2_{ri}}{4\pi^2}|\hat G_2(T_i)|^2 \\
G_{m} (K^{-1})^{m}_{n} G^{n}&=&|W|^2 + W_m(K^{-1})^{m}_{n}W^n
+(K_mW(K^{-1})^{m}_{T_1}W^{T_1}+h.c. ) \nonumber
\la{eq3}\eea
Here, $T_i, i=2,3$ and $m,n$ run over $T_1$ and the
untwisted chiral fields related to the $T_1$ sector.
Using eqs.(\ref{eq2}) and (\ref{eq3})  the scalar potential is given
by
\begin{eqnarray}
V =  e^{K} |\eta(T_2)\eta(T_3) \eta(S)|^{-4} \le(  |P|^2 \le
[\frac{S^2_r}{4\pi^2}|\hat G_2(S)|^2 +
\Sigma_i \frac{T^2_{ri}}{4\pi^2}|\hat G_2(T_i)|^2 -2 \ri] \right.
\nonumber \\
 +  \left. P_m(K^{-1})^{m}_{n}P^n
+(K_mP(K^{-1})^{m}_{T_1}P^{T_1}+h.c. )
\ri)
\la{eq4}\end{eqnarray}
where $\hat G_2(S)$ is the Eisenstein function of modular weight 2.
The potential in eq.(\ref{eq4}) is manifestly  S-duality invariant
(and $T_{2,3}$-duality invariant) since all the dependence on $S$ is
given in terms  of duality invariant functions
$e^K|\eta(S)|^{-4}\simeq (S_r|\eta(S)|^4)^{-1}$  and $S^2_r|\hat
G_2(S)|^2 $ (note that $P$ and its derivatives are $S$ and $T_i$
independent). Under quite general initial conditions  the term
$P_m(K^{-1})^{m}_{n}P^n + (K_mP(K^{-1})^{m}_{T_1}P^{T_1}+h.c. )$ will
 give a
positive vacuum potential. If $V$ is positive then, independently
of $P$ and
$P_m$, minimization of
$V_0$ with respect to $S, T_i$ gives $<S>=1, e^{-\pi/6}$ and
$<T_i>=1, e^{-\pi/6}$ (the dual invariant points)  where $S^2_r|\hat
G_2(S)|^2=
T^2_r|\hat G_2(T)|^2=0$ and $(S_r|\eta(S)|^4)^{-1}\approx 1$
take its minimum value. We will consider models where the $D$ term
vanishes and thus the $S$ dependence on $Re f$ is irrelevant.
We also assume a
$\lambda(T_1)$ which leads to a vev unity for
$T_1$ as well.

Having the
dilaton and the moduli frozen, one may then look whether inflation
occurs due to chiral fields.
In order to have inflation, the potential $V$
may not be dominated by the terms proportional  to
$P$ (implying certain cancellation conditions)
\ci{stew},\ci{am},
while to have a large number of e-folds the vev
of the inflaton field must be of the order of the Planck mass.
We look for solutions where the scale of
inflation is below the v.e.v of the higgs boson  that breaks the
gauge group to the standard model group.
Therefore, we take the inflaton
superpotenial as
$P=\Phi^2 F(\phi_i)$ where $\Phi$ is a bosonic field
that acquires a vev below the unification scale and $F(\phi_i)$
is a function of chiral fields  containing the inflaton field. The
requirements for inflation on $F$ are:  $F\ll F_i$ and $F_{ii} < F_i$
where $F_i=\partial F/\partial \phi_i$.
We also include  a $D$ term for $\Phi$. This term  gives a vev to
$\Phi$ at a scale  $M  $.  The appearance
of scalar fields with nonvanishing  v.e.vs is to be expected in
string and supergravity theories, since a symmetry breaking below
the Planck scale in a sector of the theory
introduces masses to all sectors \ci{gr}.

A simple example is given by   the superpotential
\bea
P &=& \lambda(T_1) \Phi^2 F
\nonumber \\
F&=&\phi_1^2-\phi_2^2
\la{eq6}\eea
and a $D$-term
\beq
D^2=\hat g^2(|\Phi|^2-M^2 )^2
\la{eq6a}\eeq
where $\lambda$ is the $T_1$-dependent Yukawa  coupling, $\hat g$ the
charge of $\Phi$ (which we will take as one, i.e. $\hat g=1$) and
$M $ the  gauge unification scale. As mentioned above we take
$\lambda(T_1)$ to be such that the vev of $T_1=1$ and $\lambda=1$.
This superpotential satisfies
the conditions $F\ll F_i$ and $F_{ii} < F_i$ if $\phi_1\simeq
\phi_2$. We will show that these conditions are actually dynamically
satisfied since minimization with respect to $\phi_i$ gives
$\phi_1\simeq \phi_2$. The field $\Phi$ is already at its
minimum (and cannot be used as the inflaton field) and its
vev, given in terms of $M$, will fix the magnitude of the
fluctuations.
We note that if instead of the above superpotential
a $P$ trilinear in the untwisted fields is used, it is not obvious
that  one can have
 a proper vev for $\Phi$\footnote{A
$\Phi^2$ term is needed  in order to
obtain the correct magnitude of density
fluctuations with a field
vev of the order of magnitude of  $V^{1/4}$ as expected from general
spontaneously symmetry breaking arguments.},
satisfy the cancellation conditions and achieve an end to
inflation.
The absence of trilinear terms can be explained  by imposing an
R-symmetry invariance \ci{gr}-\ci{stew1} which forbids them.

Since the vev of $\Phi$ is
much smaller than one, the normalization  factor
of the  field is not relevant. Then, in order
to simplify the presentation we consider  the $\Phi$ field to be
canonically normalized.
We take $\phi_i, i=1,2$ to be  untwisted chiral  fields belonging
to one
sector of the orbifold only (the sector associated with $T_1$).
The modular
weights of
these fields
will be different than zero only with respect to  $T_1$ and the
 K\"{a}hler potential is
\bea
K &=& K_0 + |\Phi|^2 -\ln Q
\nonumber\\
  Q &\equiv&
T_{r_{1}}-|\phi_{1}|^{2}
- |\phi_{2}|^{2}\nonumber\\
K_0&=&-ln(S+\bar{S})-ln(T_2+\bar{T_2})-ln(T_3+\bar{T_3})
\la{eq5}\eea

The
superpotential and K\"{a}hler potential of eq.(\ref{eq6}) and
(\ref{eq5})
will inflate.  The energy scale
will be set by the
value of $<\Phi>$ which is  dynamically determined  by $<\Phi> \simeq
M $.
 Demanding that the
density fluctuations coincide with the ones observed
by COBE gives the required value of $<\Phi>$. The  vev of $\Phi$
provides  for the   explanation of the smallness of the fluctuations,
since now
the factor $\lambda(T)$  is of order unity, unlike the case discussed
in
\ci{am} where the dilaton and moduli fields
are minimized due to the appearance of a gaugino
condensate and no other scale needs to be introduced.

For inflation to occur in supergravity models,
some cancelation must
occur between the terms $K' H$ and $H'$ of the derivative of the
potential, $V'=e^{K}(K'H+H')$ with $V=e^{K}\,H$,  with respect to the
inflaton field
\ci{stew},\ci{am}.  This is achieved in our example since
$\frac{\partial V}{\partial \phi_2}=0$ implies that $ <\phi_2> =
<\phi_1>$  and $P =0$.  Note that we cannot have
$G_1=K_1P+P_1=\lambda(T_1)\Phi^2(\bar\phi_1(\phi_1^2-\phi_2^2)/Q+2)=0$

and
$G_2=K_2P+P_2=\lambda(T_1)\Phi^2(\bar\phi_2(\phi_1^2-\phi_2^2)/Q-2)=0$

at the same time.   For $\phi_1=\pm \phi_2$ we have $|G_1|=|G_2|$ and
 $|G_{1,2}|_{\phi_1=\phi_2 } \leq |G_{1,2}|_{\phi_1=-\phi_2 }$. Of
course $\phi_2$ will not be identical to $\phi_1$. In particular
during inflation we know that the fluctuations of a canonically
normalized field are given by the Hawking temperature $\delta \phi =
\it{H}/2\pi$ which gives $ |<\phi_2> -<\phi_1>|  \geq \it{H}/2\pi $
(with $\it{H}$  the Hubble constant). However,
to a first approximation we can take  $ <\phi_2> = <\phi_1>$.

Using eqs.(\ref{eq4}), (\ref{eq5})  and (\ref{eq6a}) the scalar
potential is simply given by
\bea
V  &=&   \frac{e^{|\Phi|^2}}{Q}  \le( Q|G_1|^2  +Q|G_2|^2
+|G_{\Phi}|^2+2Q|G_T|^2 + Q [G_T(G_1 \phi_1+G_2 \phi_2)+h.c]
-3|W|^2\ri) +
\nonumber\\
&+&\frac{1}{2 }  (|\Phi|^2-M^2 )^2
\nonumber\\
V  &=&   \frac{e^{|\Phi|^2}}{Q} A\; \big(
Q|\frac{\bar\phi_1}{Q}P+P_1|^2+
Q|\frac{\bar\phi_2}{Q}P+P_2|^2+|P|^2 \le[ |\bar\Phi
+\frac{2}{\Phi}|^2
+2Q|-\frac{1}{Q}+a|^2-3\ri] +
\nonumber\\
&+& \le(Q \bar P(-\frac{1}{Q}+a)[(\frac{\bar\phi_1}{Q}P+P_1)\phi_1
+(\frac{\bar\phi_2}{Q}P+P_2)\phi_2] +h.c.\ri)
+\frac{1}{2 }  (|\Phi|^2-M^2 )^2
\la{eq1a}\eea
where we included the $S, T_i, i=2,3$ contributions into $A\equiv
e^{K_0}| \eta(T_2)\eta(T_3)\eta(S)|^{-4}$ and we used $a\equiv
P_{T_1}/P$, $\hat
g=1$,  $P_1=-P_2$.  Note that the potential in
eq.(\ref{eq1a}) is semi-positive definite since $ |\bar\Phi
+\frac{2}{\Phi}|^2-3 >0$.  Therefore,  the minimum is at $S=T_i=1$
giving    $A\simeq 1$ and $<Re f>=1$. Minimizing with respect to
$\phi_2$ gives $\phi_2=\phi_1$ or $P=F=0$ and eq.(\ref{eq1a}) becomes
\beq
V  =   \frac{e^{|\Phi|^2}}{Q} A\; 2 (K^{-1})_{1}^{1} |P_{1}|^2 + O(P)
+\frac{1}{2 }  (|\Phi|^2-M^2 )^2
\la{eq1b}\eeq
The derivative of the potential with respect to $\phi_1$ (using
$(K^{-1})_{1}^{1}=Q$) is
\bea
V_1=\frac{e^{|\Phi|^2}}{Q}A\;(2P_1P^1
(K_1(K^{-1})_{1}^{1}+(K^{-1})_{11}^{1}) +
(K^{-1})_{1}^{1}\bar{P^1}P_{11}) + O(P)
\la{eq2a}\eea
Note that the  term proportional to $P_1^2$ in $V_1$
cancels\footnote{Here
$K_{i}  =  \frac{\partial K}{\partial \phi_{i}}
= \frac{\bar{\phi}_{i}}{Q}$,
$K_{T_1}  =  - \frac{1}{Q}.$} (i.e.
$P_1P^1 (K_1
(K^{-1})_{1}^{1}
+(K^{-1})_{11}^{1})=0$). This cancelation can be traced to the
form of $K$ for  untwisted fields (cf. eq.(\ref{eq5})) and is
fundamental in order to have an inflationary potential.   Using
eq.(\ref{eq6}) and taking $A=1$
one can write (\ref{eq1a}) and (\ref{eq2a}) as
$
V=e^{|\Phi|^2} |\Phi|^4 \; 2|F_1|^2 +\frac{1}{2 }
(|\Phi|^2-M^2 )^2
$,
\beq
V=e^{|\Phi|^2} |\Phi|^4 \; 8|\phi_1|^2 +\frac{1}{2 }
(|\Phi|^2-M^2 )^2
\eeq
  and the derivative with respect to $\phi_1$ is
$
V_1= e^{|\Phi|^2} |\Phi|^4 \; F^1F_{11}$,
\beq
V_1= e^{|\Phi|^2} |\Phi|^4 \; 16 \bar\Phi.
\eeq
The extremum condition $V_{\Phi}=0$ gives
\beq
<|\Phi|^2>=\frac{M^2 }{1+\; 8|\phi_1|^2}
\eeq
and a scalar potential
\beq
V=\frac{8M^4 |\phi_1|^2}{1+16|\phi_1|^2}
\eeq
where we have set $e^{|\Phi|^2}=1$ since $|<\Phi>| \ll 1$. The scalar
potential $V$ vanishes at  $\phi_1=0$ and $\Phi=M $ but for
$\phi_1\neq 0$ we have  $V >0$ and the vev of $\Phi$ is $\Phi <
M $. The fact that when the inflaton potential is included
$<\Phi> $ is smaller than   $M $ ($\Phi\simeq M /3$ for
$\phi_1\simeq 1$)  can help to explain $V_{Unif}$ and $ V_{Inf}$ in
terms of a single scale (i.e. to express  $M$ in terms of the
unification scale $M_{gut}$).

The conditions for
successful inflation
 \beq
 (i)\frac{V_1}{V}  = \frac{2}{\phi_1(1+16\phi_1^2)} \ll \sqrt{K^1_1}
\hspace{1cm} (ii) \frac{V_{11}}{V}
=\frac{2(1-48\phi_1^2)}{(\phi_1(1+16\phi_1^2))^2} \ll K^1_1
\la{eqvv}\eeq
for fields with
non-canonical kinetic terms
are clearly satisfied in the region $\phi_1\simeq 1$.  Inflation
comes to an end when one of these inequalities is not satisfied. In
our example this happens first for $\frac{V_{11}}{V}=K^1_1  $ which
gives $|\phi_1|\simeq 0.65$.

The number of e-folds is given by
 \cite{am}
\bea
N &=&  -\int
\left
[\frac{V}{V'}
(K_{1}^{1}+K_{2}^{1})
\right
] d\Phi_{1} \nonumber\\
N&=& \int \frac{\phi_1 (1+16\phi_1^2)}{Q^2} d\Phi_{1}
\la{eq20b}\eea
where we have taken  the fields $\phi_i$ to be real fields,
$\sqrt{K_{1}^{1}+K_{1}^{2}} =\frac{\sqrt{2}}{Q}$ and we took the
contribution of both fields since $\phi_1=\phi_2$.
Integrating eq.(\ref{eq20b}) one finds
\beq
N= \le[ \frac{17}{4Q} + 2 log (Q/2) \ri]
\left.
\right
|_{\em init}
\la{a20}\eeq
where the subindex ``{\em init}''
refers to the initial value of the
field $\phi_1$.
{}From eq.(\ref{a20}) we see that  a large number of e-folds
requires $Q=2-2|\phi_1|^2 \ll 1$, i.e. $|\phi_1|^2 \simeq 1$. For
$Q\ll 1$ the number of e-folds can be approximated by the first term
of eq.(\ref{a20}), i.e. $N=17/4Q$.

Finally the fluctuations
are given by \ci{liddle}
\beq
\delta_H= \frac{1}{\pi\sqrt{75}} \sqrt{K_{1}^{1}+K_{1}^{2}}
\frac{V^{3/2}}{V_1}
\la{a21}\eeq
where we have  included the term
$\sqrt{K_{1}^{1}+K_{1}^{2}} =\frac{\sqrt{T_{r1}}}{Q}$ since the
inflaton is not canonically normalized. Using eqs.(\ref{eqvv}),
(\ref{a20}) (\ref{a21})
for $Q\ll1$ the fluctuations can be written as
\beq
\delta_H=\frac{4N}{\pi\sqrt{150}}\;V^{1/2}
\la{eq21b}
\eeq
The field values where
we calculate the fluctuations are determined
by the horizon scale today:
for a fluctuation emitted with a certain wavelength
during inflation, one may
calculate the wavelength
that the fluctuation has
today and compare this to the horizon
distance $(6000 \; {\rm Mpc})$
\cite{fluctu}. Indeed, during inflation
a wave emitted at some
value $\phi_{1}$
increases its wavelength. In order to solve the horizon and flatness
problems
the number of e-folds of inflation  is  \ci{liddle}
\beq
N=62 +ln\le(\frac{V_{Inf}^{1/4}}{10^{16}GeV}\ri)
+ln\le(\frac{V_{Inf}^{1/4}}{V_{end}^{1/4}}\ri)
-\frac{1}{3}ln\le(\frac{V_{Inf}^{1/4}}{\rho_{reheat}^{1/4}}\ri)
\la{N}\eeq
where $\rho_{reheat}= \frac{\pi^2g_*(T_R)}{30} \;T^4_R $ is the
energy density after reheating, $g_*$ the number of relativistic
degrees of freedom (915/4 for the minimal supersymmetric standard
model) and
$T_R$ the reheating temperature.
The exact time of emission
of a fluctuation with horizon
size today, depends on the value of
$T_{R}$, which is model dependent.
To determine $T_R$ we calculate  the width of the inflaton field
$\phi_1$. This  field will decay into states $\Xi$ of another sector
of the theory by the interaction $\partial V/\partial \phi_1W(\Xi)$
where the superpotential W should be replaced by the scalar
components \ci{gr}.  This interaction generates a trilinear coupling
to the light fields of strength $\sim m_{\phi_1}$ giving a width
$\Gamma_{\phi_1} \sim m^3_{\phi_{1}}/(2\pi)^3$, where $m_{\phi_1}$ is
the mass of the inflaton. The temperature associated with the
radiation of the inflaton decay is \ci{gr}
\beq
T_R \sim \le( \frac{30}{\pi^2g_*}\ri)^{1/4}
\Gamma_{\phi_{1}}^{1/2}\simeq 2.2\times 10^{-2} m_{\phi_1}^{3/2}
\la{tr}\eeq
and the mass of the inflaton is calculated at the end of inflation.

Solving eqs.(\ref{N}),  (\ref{tr}) and  (\ref{eq21b})
for $\delta_H = 2.5 \times 10^{-5}$,  as required by COBE \ci{COBE}
we obtain
$N\simeq 56$, $T_R=4.5\times 10^8 GeV, m_{\phi_1}= 1\times 10^{13}
GeV$ and
an inflation  scale  of $V^{1/4}\simeq 5\times 10^{15}$ GeV. Note
that
the reheating temperature is consistent with the bounds put by the
relic abundance if the gravitino mass if of the order $1 TeV$
\ci{gr},\ci{relic}.

The spectrum of fluctuations in this example is almost scale
invariant. The spectral index is \ci{liddle}
\beq
n=1+2
\frac{1}{K_{1}^{1}+K_{1}^{2}}
\frac{V_{11}}{V}-\frac{1}{3(K_{1}^{1}+K_{
1}^{2})}\le(\frac{V_1}{V}\ri)^2= 0.99
\eeq
where we have again taken into account for the non canonical kinetic
term of $\phi_1$ and included the contribution of $\phi_2$ since
$\phi_2=\phi_1$. This spectral index corresponds to a slightly tilted
spectrum which has less power on galactic scales in a cold dark
matter universe and  therefore agrees better with the observations.

We have seen that a simple superstring model
can lead to enough e-folds of inflation to solve the horizon and
flatness problems and the density fluctuations of the inflaton field
can be normalized to COBE once the dilaton field has been stabilized.

The stability of $S$  is model independent (i.e. does not
dependent on the chiral superpotential $P$) and
 allows to have an inflationary potential  at any scale. Note that
the
only condition needed is to have a positive potential but this is
clearly no constraint since any inflationary potential must be
positive anyway.  Therefore, with an $S$-duality potential there is
no longer a need  to be concerned about the dynamics of the dilaton
field and one can look for inflationary potentials in  general
supergravity models.

It is also interesting to note that with an
S-dual potential we can have
two or more stages of inflation.  The first may occur
at a large scale, as we have described so far, and  depends on
the scale of a spontaneously broken  symmetry.  This stage of
inflation will solve the horizon and flatness problems and will give
rise to the density fluctuations observed by COBE.
 An additional stage of
inflation may occur
below the supersymmetry breaking scale \ci{am}. This later scale of
inflation is welcomed to solve the Polonyi problem \ci{Randall}.  Of
course, if the number of e-folds of inflation at the second stage is
very large, then it will erase the
original density fluctuations generated at the first stage and the
resulting
density fluctuation will be  too small.
Providing that this does not occur,
even if we assume that there exist two stages of inflation,
the spectrum of
fluctuations that we predict is approximately a scale-invariant
Harrison-Zeldovich one, and comes from the
first stage of inflation. This differs from
other models which use
two stages of inflation \cite{twoi} so as to reconcile the
observed discrepancy
between the COBE observations and
the existing cold or hot dark matter
models for structure formation
\cite{p3}.
However, this discrepancy may be explained,
either by considering schemes where both
cold and hot dark matter are present,
or by taking
into account the effect of additional
sources of fluctuations.
For example, we have shown
that, unlike what was previously thought,
under certain conditions,
domain walls may enhance the standard
cold dark matter spectrum without
inducing unacceptable cosmic microwave
background distortions.
This occurs provided that either
one of the minima of the potential of the scalar
field $\phi$ is favoured \cite{wa2}, or
the domain walls are unstable and annihilate
after having induced fluctuations to the
cold dark matter background \cite{wa1}.
Such solutions are predicted to exist in superstring
models, therefore we believe that the overall
picture that we have for inflation and structure formation
in the framework of these theories is consistent.

Finally, let us comment on the possibility
that for a fixed $P(\phi), P_m(\phi)$
the dilaton and/or moduli can  work as inflaton fields.
For $S$ far away from its minimum ($S\gg1$), the potential has  an
exponential behaviour  due to the $\eta(S)$ term, i.e. $V_0 \simeq
e^{\alpha S_r}$, with coefficient $\alpha=\pi/12$.
Since this potential is not flat enough
it does not lead to inflation. The same
conclusion holds for the moduli fields $T_{2,3}$. Around the
extremum $S=1, e^{-\pi/6}$ the potential is flat enough, however
inflation will not come  to an end if $P, P_m$ are constant
, since in this case the potential will remain
positive, even at the extremum of $S$. It
is therefore unnatural to assume that $P(\phi), P_m(\phi)$ do not
vary and we have to study their dynamics.
In this framework, suppose first that
 $\phi$ has a local minimum; then it is possible to
find an "old" inflationary
potential where the general minimum is obtained by vacuum tunneling
from the local minimum. However,   it is well know that such a
solution leads to phenomenological problems
(unless going to models of extended inflation).
On the other hand, we can examine the case with a
continuous variation w.r.t. $\phi$. In this case, since the
 variation of $\phi$ will in
general be larger
 than that   of $S$ around its extremum,
the inflaton field will then correspond to $\phi$ and not to
$S$.

\vspace{0.4 cm}

\noindent
{\large {\bf Acknowledgment }}

\noindent
We would like to thank  G. G. Ross for
suggesting S-duality as a mechanism of
freezing the dilaton, and for
many enlightening discussions and comments
on inflationary models.


\begin{thebibliography}{99}

\small{

\bibitem{inflation}
E. B. Gliner, JETP 22, 378 (1966);
D. Kazanas, Ap. J. 241, L59 (1980);
A. H. Guth, Phys. Rev. D23, 347 (1981);
G. D. Coughlan, R. Holtman, P. Ramond and G. G. Ross,
Phys. Lett. B140, 44 (1984).

\bibitem{faults}
For a review of the open questions of the
standard hot big-bang theory see
E. W. Kolb and M. S. Turner
{\em ``The Early Universe''}, Frontiers in Physics 69
(1990), chapter 8 and references therein.

\bibitem{chao}
A. D. Linde, Phys. Lett. B129, 177 (1983).

\bibitem{wett}
Q. Shafi and C. Wetterich,
Phys. Lett. B152 (1985) 51;
Nucl. Phys. B289 (1987) 787;
Nucl. Phys. B297 (1988) 697;
C. Wetterich, Nucl. Phys. B324 (1989) 141.

\bibitem{stew}
B. A. Campbell, A. Linde and K. Olive, Nucl. Phys. B355 (1991) 146;
N. Kaloper
and K. A. Olive,
Astropart. Phys. 1 (1993) 185;
 E. J. Copeland, A. R. Liddle ,  H. Goldberg, NUB-3082-93-TH preprint
hep-ph/9312282;  M. Bento and O. Bertolami, gr-qc/9505038 and
gr-qc/9507045.

\bibitem{sdualcosmo}
E.I. Guendelman Gen. Relativ. Gravit. 23 (1991) 521; Supriya Kar,
Hrvendra Singh  IP-BBSR-95-66, hep-ph/9507063


\bibitem{gr}
G. G. Ross and S. Sarkar, hep-ph/9506283 preprint.

\bibitem{stew1}
D. H. Lyth, E. D. Stewart, D. Wands,
Phys.Rev.D49 (1994) 6410;
E. D. Stewart, Phys. Rev. D51 (1995) 6847

\bibitem{Venez} M. Gasparini and G. Veneziano, Phys. Rev. D50 (1994)
2519.

\bibitem{am}
A. De la Macorra and S.Lola,
 ``Inflation from superstrings''
IFUNAM-FT-94-63, HD-THEP-94-45,
hep-ph/9411443

\bibitem{stein}
R. Brustein and P. J. Steinhardt, Phys. Lett.
B302 (1993) 196;  P. Binetruy and M. K. Gaillard, Phys. Rev. D34
(1986) 3069; B. de Carlos et al., Phys. Lett. B318 (1993) 447.

\bibitem{gaug}
For a review see D. Amati, K. Konishi, Y. Meurice,
G. Rossi and G. Veneziano, Phys. Rep 162 (1988) 169.

\bibitem{axel}
A. De La Macorra and G. G. Ross,
Nucl. Phys. B404 (1993) 321;
Phys. Lett. B325 (1994) 85.

\bibitem{COBE}
G. F. Smoot et al., Astrophys. J. Lett.
396 (1992), L1.


\bibitem{Sdual}
A. Font, L. E. Ibanez, D. L\"{u}st and
F. Quevedo, Phys. Lett. B249 (1990) 35.


\bibitem{seibwit}
N. Seiberg and E. Witten,
Nucl. Phys. B426 (1994) 19;
Nucl. Phys. B431 (1994) 484.

\bibitem {nilles} J. Horn and G. Moore, Nucl. Phys. B432 (1994)109;
Z. Lalak, A. Niemeyer, H.P. Nilles
Phys.Lett.B349 (1995) 99; Z. Lalak, A. Niemeyer, H.P. Nilles
hep-ph/9503170; Pierre Binetruy, Mary K. Gaillard. LBL-37198,
hep-th/9506207

\bibitem{r15}E. Cremmer, S. Ferrara, L. Girardello and
          A. Van Proeyen, Nucl.Phys. B212 (1983) 413.


\bibitem{effective}
For a review of 4D string theories see
M. Green, J. Schwarz and E. Witten,
Superstring Theory, Cambridge University
Press, 1987.

\bibitem{Bellido} See for instance L.J. Garay and J. Garcia-Bellido,
Nucl. Phys. B400 (1993) 416.


\bibitem{fluctu}
A. Linde,
{\em ``Particle Physics and Inflationary Cosmology''}, Harwood
Academic,
Switzerland (1990)
and references therein.



\bibitem{liddle}
A. R. Liddle and D. H. Lyth, Phys. Rep. 231 (1993) 1

\bibitem{relic}
J. Ellis, D.V. Nanopoulos and S Sarkar, Nucl. Phys. B259 (1985) 175
J. Ellis, G.B. Gelmini and S. Sarkar, Nucl. Phys. B373 (1992) 399; S.
Sarkar, Oxford prerpint OUTP-95-16p;  M. Kawasaki and T. Moroi, Prog.
Theor. Phys.93 (1995)879
\bibitem{twoi}
D. Polarski and A. Starobinski, Nucl. Phys. B385 (1992) 623;
P. Peter, D. Polarski and A. Starobinski, Phys. Rev. D50 (1994) 4827.

\bibitem{p3}
M. Davis, F. J. Summers, D. Schlegel, Nature 359, 393 (1992);
A. N. Taylor, M. Rowan-Robinson, Nature 359, 396 (1992);
W. Saunders, M. Rowan-Robinson, A. Lawrence, Mon. Not.
R. Astr. Soc. 258, 134 (1992).

\bibitem{wa2}
Z. Lalak, S. Lola, B. A. Ovrut and G. G. Ross,
Nucl. Phys. B434 (1995) 675.

\bibitem{wa1}
S. Lola and  G. G. Ross, Nucl. Phys. B406 (1993) 452.

\bibitem{Randall}
L.Randall and S.Thomas, NP 449B (1995) 229;
T. Banks, M. Berkooz and P.J. Steinhardt, Phys. Rev. D52
(1995) 705.




}

\end{thebibliography}
\end{document}